\documentclass[AMA,STIX1COL]{WileyNJD-v2}

\articletype{Research Article}%

\received{26 April 2016}
\revised{6 June 2016}
\accepted{6 June 2016}

\usepackage[utf8]{inputenc} 
\usepackage[T1]{fontenc}    
\usepackage{hyperref}       
\usepackage{url}            
\usepackage{booktabs}       
\usepackage{amsfonts}       
\usepackage{nicefrac}       
\usepackage{microtype}      
\usepackage{lipsum}
\usepackage{amsmath}
\usepackage{graphicx}

\usepackage{xcolor}

\begin{document}

\title{A statistical model to assess risk for supporting SARS-CoV-2 quarantine decisions}


\author[1]{Sonja Jäckle*}

\author[2]{Elias Röger}

\author[3]{Volker Dicken}

\author[3]{Benjamin Geisler}

\author[4]{Jakob Schumacher}

\author[3]{Max Westphal}

\authormark{Jäckle \textsc{et al.}}

\address[1]{\orgname{Fraunhofer MEVIS - Institute for Digital Medicine}, \orgaddress{\state{Lübeck}, \country{Germany}}}

\address[2]{\orgname{Fraunhofer ITWM - Institute for Industrial Mathematics}, \orgaddress{\state{Kaiserslautern}, \country{Germany}}}

\address[3]{\orgname{Fraunhofer MEVIS - Institute for Digital Medicine}, \orgaddress{\state{Bremen}, \country{Germany}}}

\address[4]{\orgname{Health Department Berlin-Reinickendorf}, \orgaddress{\state{Berlin}, \country{Germany}}}

\corres{*Sonja Jäckle, Fraunhofer MEVIS - Institute for Digital Medicine, Maria-Goeppert-Straße 3, 23562 Lübeck, \email{sonja.jaeckle@mevis.fraunhofer.de}}



\abstract[Abstract]{In February 2020, the first human infection with SARS-CoV-2 was reported in Germany. Since then, the local public health departments have been responsible to monitor and react to the dynamics of the pandemic. One of their major tasks is to monitor infected persons. When they have attended group meetings e.g. at school, at a sports event, the contacts have to be traced and potentially are quarantined for a period of time. When all relevant contact persons obtain a negative polymerase chain reaction (PCR) test result, the quarantine may be stopped. However, tracing and testing of all contacts is time-consuming, costly and (thus) not always feasible.

This motivates our work, in which we present a statistical model for the probability that no transmission of SARS-CoV-2 occurred given an arbitrary number of test results at potentially different time points. Hereby, the time-dependent sensitivity and specificity of the conducted PCR test are taken in account. We employ a parametric Bayesian model which can be adopted to different situations when specific prior knowledge is available. This is illustrated for group events in German school classes and applied to exemplary real-world data from this context.
Our approach has the potential to support important quarantine decisions with the goal to achieve a better balance between necessary containment of the pandemic and preservation of social and economic life. The focus of future work should be on further refinement and evaluation of quarantine decisions based on our statistical model.}


\jnlcitation{\cname{%
\author{Jäckle S.}, 
\author{Röger E.}, 
\author{Dicken V.}, 
\author{Geisler B.}, 
\author{Schumacher J.} and 
\author{Westphal M.}} (\cyear{2021}), 
\ctitle{A statistical model to assess risk for supporting SARS-CoV-2 quarantine decisions}, \cjournal{arXiv preprint arXiv:2010.15677}, \cvol{2020}.}

\maketitle

\section{Introduction}

\subsection{SARS-CoV-2 pandemic and situation in Germany}
The first cases of human SARS-CoV-2 infections were reported in China. Subsequently, this virus spread to almost all countries in the world, causing most countries to be seriously affected~\cite{whoNumbers}. Many countries legislated strict lock-downs in March 2020 resulting in a decreasing number of new infections. Afterwards, more liberal rules applied such as social distancing, limits on larger gatherings and rules to use face masks. In addition, many countries have issued regulations requiring infected persons to be isolated and contact persons to be quarantined. 
In Germany, the local health departments report information about local new cases on a daily basis to the Robert Koch Institute (RKI), the central German institution for infectious disease study and control. The RKI reports on a regular basis about the SARS-CoV-2 situation in Germany, see for example~\cite{rki2020Fallzahlen}. Based on those reported numbers, the RKI recommends an implementation, adaption or cancellation of procedural rules for political decision-makers in Germany. 

Since August 2020, a second significant increase of new infections per week has been reported in Germany as well as in many other European countries. Many of these cases were associated with vacation travels. In Germany, it resulted in several times higher numbers than reported in June, where the number of new infections per day had reached a minimum so far~\cite{rki2020Fallzahlenanstieg}. This triggered strong societal discussions about risks involving the reopening of schools and re-allowing events such as mass gatherings. Moreover, a shortening of the quarantine duration time for contact persons has been discussed. This infection increase resulted in the so-called second wave in end of 2020 and a third wave of infections also occurred in spring of 2021.

\subsection{Quarantine decisions in Germany}

The German Protection against Infection Act ("Infektionsschutzgesetz")~\cite{IfSG2020} defines the general rules of dealing with infectious diseases. It specifies which pathogens must be monitored and allows German health departments to quarantine people at home. In the current SARS-CoV-2 pandemic, all contact persons of an infected persons with a high infection risk had to stay at home. Since a large number of SARS-CoV-2 cases was reported by the RKI, a large number of contact persons probably had to be quarantined in Germany~\cite{rki2020Fallzahlen}. The RKI provides the procedural guidelines for theses cases. In the end, however, the local health departments have to decide autonomously if and how long a person or a group of persons have to be quarantined. 

For the German health departments, this is the first time in many decades where they have to quarantine such a large number of infected persons and contact persons. In the last decades, domestic quarantine has been imposed only a few times. For example in 2002/2003, the first severe acute respiratory syndrome (SARS-CoV) started to spread over several countries. In Germany, contact persons were quarantined at home, but the number of quarantined was substantially smaller than in the current SARS-CoV-2 pandemic~\cite{rki2006PublikationSARS}. A default workflow for contact tracing of SARS-CoV-2 cases is provided by the RKI~\cite{rki2020Kontaktpersonen}. These contact persons are classified into one of the following three categories:

\begin{itemize}
	\item category 1 with close contact (high infection risk)
	\item category 2 (low infection risk)
	\item category 3 (medical staff with adequate protection)
\end{itemize}

The RKI defines category 1 as persons who  
\begin{itemize}
	\item had face-to-face contact for at least 15 minutes or
	\item had direct contact to bodily fluids or 
	\item were exposed to a relevant aerosol concentration.
\end{itemize}

Other contact persons are classified as category 2. Medical staff with appropriate personal protective equipment is classified as category 3. The RKI recommends that contact persons of category 1 and 2 normally should reduce their contacts to other persons, and contact persons of category 1 additionally should stay in home quarantine until 14 days after exposure. Moreover, contact persons of category 1 should be tested, ideally one day after the case documentation and a second time 5-7 days after the contact with the SARS-CoV-2 case~\cite{rki2020Kontaktpersonen}. 

Since February 2020, a high number of contact persons of SARS-CoV-2 had to stay in home quarantine, especially in March and April 2020, when the highest number of active SARS-CoV-2 cases occurred. The RKI collected data about all SARS-CoV-2 spreads until August $11^{\text{th}}$ 2020 and reported the frequency and size for different event types~\cite{rki2020Infektionsumfeld}. The probability of a transmission and the number of infected people depends noticeably on the event type. As of August $11^{\text{th}}$ 2020, 3902 spreads occurred in households and 709 spreads in retirements homes, whereas only 31 spreads were reported at schools and 33 at kindergartens~\cite{rki2020Infektionsumfeld}.

This indicates that the risk of a SARS-CoV-2 spread in a group depends highly on the scenario. This fact might inform quarantine decisions and might justify more liberal quarantine decisions in certain situations. However, the event type is currently not taken into account systematically according to the RKI recommendations. It is often not possible to get test results from all participants of a contact event. Reasons for this might be that the test capacity is limited, the persons could not be contacted at all or they did not go to a test center. 

Moreover, in the overall view quarantine can indirectly cause economic damage. Since the persons have to stay at home or employees have to take care of their quarantined children, they are unable to work for the whole time of the quarantine. In addition, quarantines of persons, which work in system relevant positions, can lead to an undersupply of critical infrastructure. This includes medical care and food supply, but also other sectors, such as government administration and supply of water and energy. 

Against this background, it is desirable to support these quarantine decisions with objective data and statistical models, because many people are involved in the process. A software implementing such a model can streamline and support the decision making process and make it less prone to human mistakes. To the best of our knowledge, this is the first work which describes a statistical model as decision support for group quarantines.

\subsection{Structure of the paper}

In this work, we will provide a statistical model for the risk assessment of canceling group quarantines. It allows to estimate the risk of canceling a group quarantine, where a fraction of the group has been tested and all test results are negative. In section~\ref{section:methods}, the Bayesian model including the prior choice is described. It is applied for a school class scenario and analyzed in section~\ref{section:results}. Finally, the model and the results are discussed in section~\ref{section:discussions}.

\section{Materials and Methods}
\label{section:methods}

\subsection{Use case}

In the following, we describe an idealized use case for our model. We assume that a group contact event took place, e.g. a family party, a sports event or a teaching unit in school.
A certain time after the event, one or multiple participants are tested positive for SARS-CoV-2, possibly only after the onset of symptoms. More generally, a series of contact events with the same group (e.g. in school or at work) might have taken place. 
We presume that there is only a single person known to be infected, the so-called \textit{primary} or \textit{index case}. Adaptations for multiple primary cases are discussed in section \ref{section:discussions}. The remaining group, without counting the primary case, is assumed to consist of $M$ individuals with unknown infection status. As a precautionary measure, this group of potential \textit{secondary cases} is quarantined by health authorities to contain further infection dynamics. In addition, all contact persons are invited to take a polymerase chain reaction (PCR) test. Currently, all contacts should stay in quarantine for 14 days available~\cite{rki2020Quarantaene}. 

\begin{figure}[ht]
	\centering
	\includegraphics[width=0.59\textwidth]{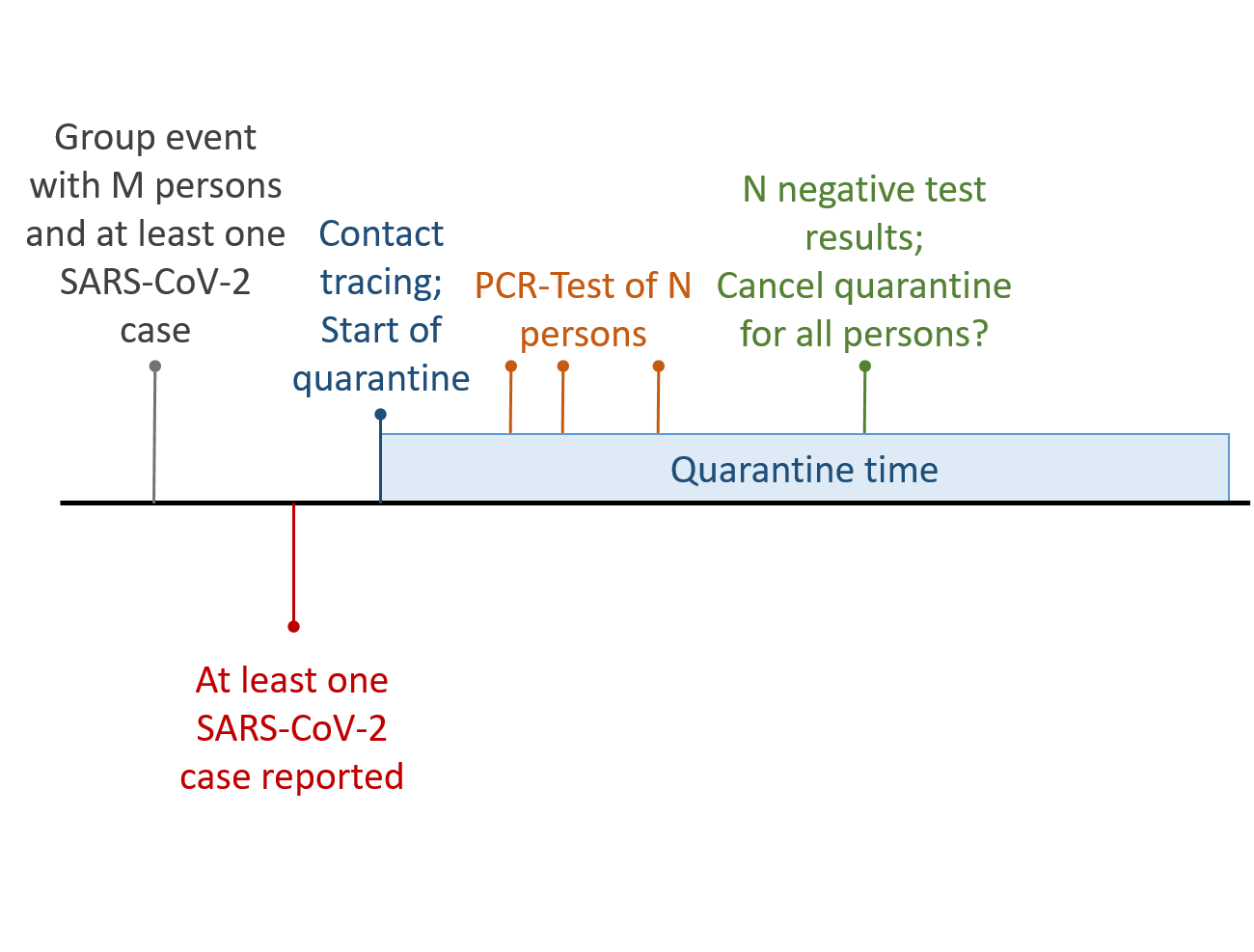} 
	\includegraphics[width=0.39\textwidth]{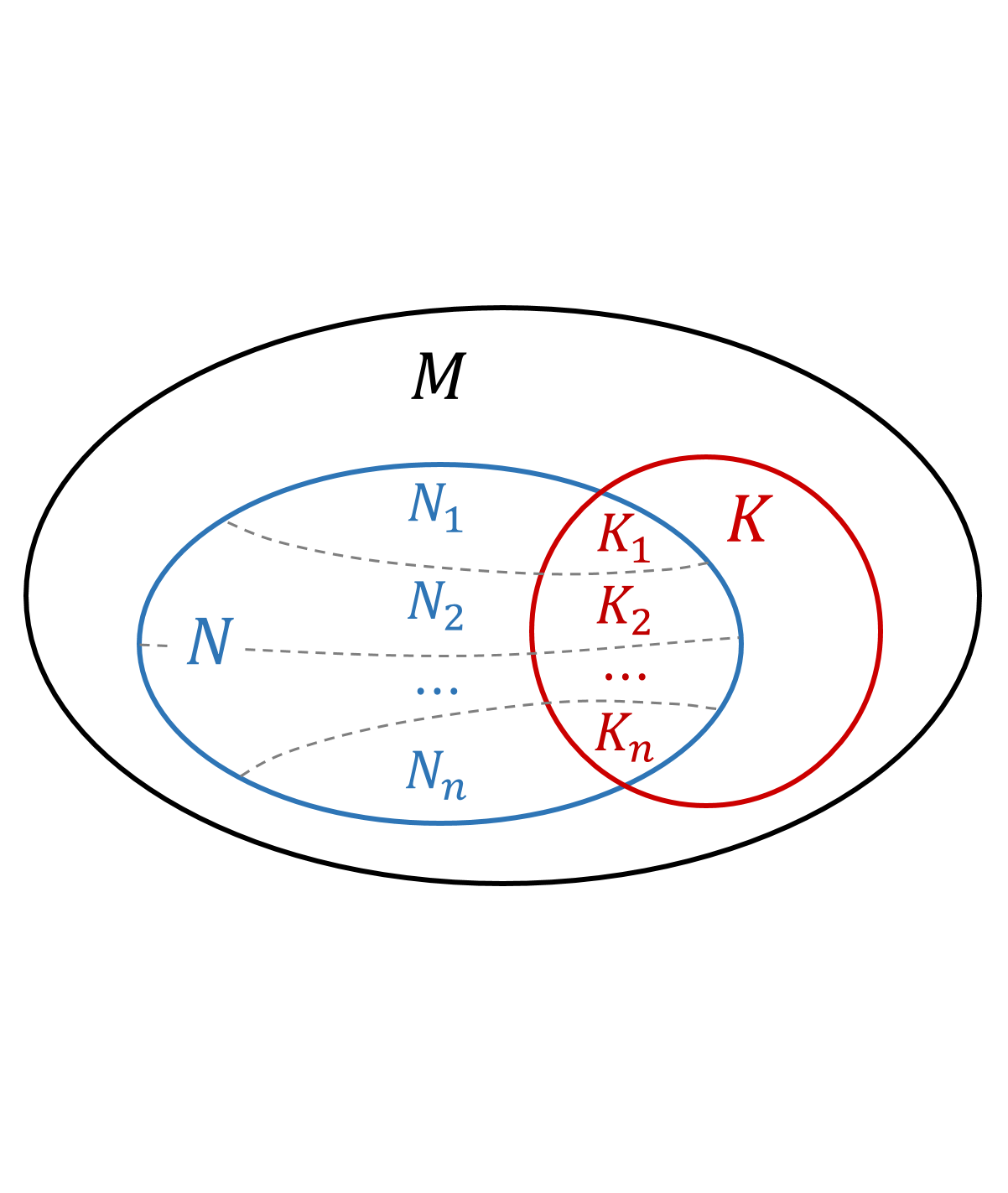}
	\caption{Left: Timeline of the events starting from the group event until the quarantine decision. Right: Overview of the group with $M$, the possible number of infected $K$ and the negative tested persons $N$. For $N$ and $K$ the contact persons are divided into subgroups $N_d$ and $K_d$, which are tested on day $D_d, d \in \{1,...,n\}$ are also illustrated.}
	\label{fig:illu}
\end{figure}

We are now interested whether there occurred any transmissions of SARS-CoV-2 from the primary case to any of the contact persons during the group event. Any other transmission paths outside of the group event are not considered. 
If there is at least one positive test result in the remaining group, this indicates that a SARS-CoV-2 transmission occurred. In this case, the health office in Germany are not allowed to cancel the group quarantine, because other persons might be also infected and this has to be checked. Thus, the case of interest is that all test results are negative. Figure \ref{fig:illu} (left) illustrates the timeline of these events described above. The main question which guided the methodological developments in this work is:

\begin{quote}
	"Under which conditions can a group quarantine be released (earlier) such that the probability of overlooking a (secondary) infection is low?"
\end{quote}

In an ideal world, all potential cases would be tested by a highly accurate test. However, we rather suppose that not all $M$ individuals are tested due to resource constraints or other reasons. Instead, out of the $M$ person group $N \leq M$ are successfully tested negative. 

In our situation, we consider reverse transcription PCR tests with (time-dependent) imperfect accuracy. An extension to other tests, as for example the rapid antigen test, is possible as long as the specificity and sensitivity can be characterized. Thus, the probability that all test results are negative even though there are some infected persons in the group is not zero. As a further complication, we allow that not all $N$ tests are conducted on the same day (after the group contact event). Rather, usually test time points vary due to logistic or resource limitations. In summary, we have only partial (probabilistic) knowledge regarding the disease status of the remaining group. The above question thus can be formulated more precisely as:

\begin{quote}
	"How likely is it that there is no secondary infection, given all $N$ out of $M$ people are tested negative?".
\end{quote}

The model described in the following assumes that the people to be tested are chosen randomly. This is not likely to hold in the real world as most authorities will give priority to testing people that were close spatially or socially to the primary case(s). As a result, the risk of an transmission will be overestimated and the probability that no infection occurred will be lower. Thus, our model will give a conservative estimation. In consultation with the health office Berlin-Reinickendorf, we concluded that this risk averse calculation is appropriate.

\subsection{Statistical model} 
\label{subsec:model}
The goal of the model is to estimate the probability, that no SARS-CoV-2 transmission occurred. As outlined above, we consider $M$ contact persons, $N \leq M$ of them having been tested and all tests results are negative. Moreover, we have an unknown number of persons $K; K \leq M$, which were actually at the group event. For this purpose, we consider only the secondary transmissions caused by group event, but no further potentially infections outside of the participant group $M$. Figure~\ref{fig:illu} (right) gives an overview over these (sub)groups. According to Bayes rule the posterior distribution for a given number of $K = k$ is then described as

\begin{align}
P(K = k | \text{N out of M tested negative}) &= \frac{P(\text{N out of M tested negative}| K=k) \cdot P(K = k)}{P(\text{N out of M tested negative})}. 
\end{align}  

For our use case, we are interested in the probability, that no one is actually infected ($K=0$) given that $N$ out of $M$ have been tested and all tests result are negative. This probability is then given as

\begin{align}
p_0 &:= P(K = 0 | \text{N out of M tested negative}) \\ 
&= \frac{P(\text{N out of M tested negative}| K=0) \cdot P(K = 0)}{P(\text{N out of M tested negative})} \\
&= \frac{P(K=0)}{\sum_{i=0}^{M} P(\text{N out of M tested negative}|K=i)P(K=i)} \label{equ:Bayes},
\end{align}  
since it is assumed, that healthy persons are always tested negative, which is described in more detail in section~\ref{subsec:likeli}. Thus, $P(\text{N out of M tested negative}| K=0) = 1$ holds.

This requires an a-priori probability distribution $P(K)$ for the number of infected persons $K$. This prior describes which values of $K$ are considered probable before considering the data. Furthermore, a likelihood function $P(\text{N out of M tested negative}|K)$ is needed, which describes the distribution probability that $N$ of $M$ are tested negative under the condition that $K$ persons are infected in the group. The modeling of the prior and the likelihood function are described in the following sections. 

\subsection{Prior distribution}
\label{modeling:prior}

Transmissions of SARS-CoV-2 have a high heterogeneity, which means that most infected do not infect anybody else whereas some infected persons are causing super-spreading events. Lelieveld et al.~\cite{aerosol2020} introduced a model based on the binomial distribution for SARS-CoV-19 aerosol transmission risk calculation. 
We employ a Beta-binomial prior instead, which is a direct extension to the binomial prior. The uncertainty regarding the transmission probability $p$ can be modeled using the Beta-binomial as prior distribution. 

The Beta-binomial distribution is defined as

\begin{equation}
K \ | \ \alpha, \beta, M  \sim \text{BB}(M, \alpha, \beta).  \label{equ:BB}
\end{equation} 

with shape parameters $\alpha$ and $\beta$. It is a discrete probability distribution on a finite support of non-negative integers and it is frequently used when the success probability in each of the known number of trials is either unknown or random. Its probability mass function is defined as

\begin{equation}
P(K=k) = \binom{M}{k} \frac{\text{B}(k+\alpha,M-k+\beta)}{\text{B}(\alpha,\beta)} \label{equ:PMF}
\end{equation} 
where $\text{B}(\alpha,\beta)$ is the beta function. 
The probability that the contact persons become infected and the number of infected persons depends on the type of event, e.g., if it takes place outside or inside and how much the persons interact and talk. Thus, the parameters of the Beta-binomial distribution should be adjusted to the considered use case.

In order to define $\alpha$ and $\beta$, we use two other parameters for the scenario specific modeling which allows for easier interpretation. We consider the probability that a transmission of SARS-CoV-2 to at least one person takes place

\begin{equation}
P(K>0)
\end{equation}

and the conditional expected value

\begin{equation}
E(K|K>0) = \frac{\sum_{i=1}^{\infty}P(K=i)\cdot i}{P(K>0)}.
\end{equation}

With these two known values, the needed parameter $\alpha,\beta$ can be determined by solving an optimization problem, for example, with the L-BFGS method~\cite{byrd1995limited}.

\subsection{Likelihood}
\label{subsec:likeli}

The accuracy of a test is described by the specificity (Sp, also called \textit{True Negative Rate}) and the sensitivity (Se, also called \textit{True Positive Rate}). In this paper, PCR tests are considered for testing. The U.S. Food and Drug Administration analyzed PCR tests and determined a specificity of nearly
$100\%$~\cite{pcrTestspec2020} (the very few false positives are usually caused by laboratory handling errors). Thus, we assume a the specificity of 1. However, the sensitivity is less than $1$ and the probability that a person infected with SARS-CoV-2 is tested positive depends on the time point of testing, see also~\cite{kucirka2020variation}. For our model, these probabilities were read and copied manually from the first graph of Fig. 2 of publication~\cite{kucirka2020variation}. Thus, the test of an infected person will be negative with almost $100\%$ probability on the first and second day after the infection; between days six to ten after infection this probability of a false negative test result drops below 25\%, reaching a minimum of about 20\% on day eight.

This day specific sensitivity is used for our realistic modeling of the likelihood function:
The $N$ tested persons are split in $n$ day specific groups. $N_1$ persons were tested on day $D_1$ after the group event, $N_2$ persons were tested on day $D_2, D_2 \neq D_1$, ... and $N_n$ persons were tested on day $D_n, D_n \neq D_d \text{ for } d \in \{1,...,n-1\}$. In the same way, the $K$ infected persons are divided into days: $K_1, ..., K_n$. For both the tested persons $N_d$ and the infected tested persons $K_d$, we assume that they are only tested on a single day. Thus, so every person $x \text{ is tested on day } D_i \text{ and is not tested on day } D_j \text{ and } i \neq j \; \forall \: i,j \in \{1,...,n\}$. The remaining untested and infected persons $N_0 = M - N, K_0 = K - \sum_{d=1}^{n}K_d$ are considered as an additional group. These subsets of $N$ and $K$ are also illustrated in the right image of Figure~\ref{fig:illu}. Moreover, the day dependent sensitivity $s_{D_d}$ is known for all days $D_d, d \in \{1,..., n\}$.

The distribution of SARS-CoV-2 infected persons $K$ can now be considered as a random drawing from an urn containing $M$ persons divided into the subsets $N_0, N_1,..., N_n$. Then, the probability of one specific draw $\Tilde{K} = (K_0,...,K_n), \sum_{d=0}^{n}K_d = K \text{ and } K_d \leq N_d \;\; \forall \: d \in \{0,...,n\}$ can be calculated with the formula of the multivariate hyper-geometric distribution:

\begin{align} 
P(\Tilde{K} = (K_0,...,K_n)) = \frac{\prod_{d=0}^{n}\binom{N_d}{K_d}}{\binom{M}{K}} 
\label{equ:Hypergeom}
\end{align}

Furthermore, the probability for a negative test one day $d$ is needed. For healthy persons, this probability is 1, since we assume a specificity of $100\%$. But for infected persons, this probability depends on the test date:

\begin{align} 
&P(\text{Negative test on day } D_d | \text{ person is healthy}) = 1;  \\
&P(\text{Negative test on day } D_d | \text{ person is infected}) =  1-s_{D_d}
\end{align}

On each day $D_d$, $N_d$ persons with an unknown number of infected persons $K_d \leq N_d$ are tested and the likelihood of testing them all negative is given by 

\begin{align} 
P(N_d \text{ negative tests on day } D_d | K_d \text{ persons are infected}) &= (1-s_{D_d})^{K_d} \cdot 1^{N_d-K_d} \\
&= (1-s_{D_d})^{K_d}
\end{align} 

The probability for testing all infected persons on all days negative can be determined by multiplying all day specific probabilities:

\begin{align} 
&P(N \text{ negative tests overall}| K = \sum_{d=0}^{n}K_d \text{ of }N_d \text{ persons  are infected}) \\ 
&= \prod_{d=1}^n P(N_d \text{ negative tests on day } D_d | K_d \text{ of }N_d \text{ persons are infected}) \\
&= \prod_{d=1}^n (1-s_{D_d})^{K_d}. 
\label{equ:negativTest}
\end{align}

For the next step, we are considering a simple example, where all persons are tested on the same day and it holds $N = N_1, N_0 = M - N$. In this case, the following draws $\Tilde{K}$ can occur:

\begin{align}
X_K = \{ (\min(N_0,K), K - \min(N_0,K)), \dots , (K - \min(N_1,K), \min(N_1,K)) \}
\end{align}

Thus, the probabilities for all those possible draws defined in equation~\eqref{equ:Hypergeom} multiplied with the negative testing probabilities defined in equation~\eqref{equ:negativTest} have to summed up to obtain

\begin{align}
P(\text{N out of M tested negative}|K) = \sum_{(K_0,K_1)\in X_K}\frac{\binom{N}{K_0}\binom{M-N}{K_1}}{\binom{M}{K}}(1-s_{D_1})^{K_1}.
\end{align}

In general, the persons are tested on $n$ different days and we obtain:

\begin{equation} 
X_K = \left\{(K_0,...,K_n) | \sum_{d=0}^{n}K_d = K \text{ and } K_d \leq N_d \;\; \forall \: d \in \{0,...,n\}  \right\}
\end{equation}

as set of all possible draws. Again the probabilities have to summed obtain the probability 

\begin{equation} 
P(\text{N out of M tested negative}|K) = \sum_{\Tilde{K} \:\in X_K} P\left(\Tilde{K} = (K_0,...,K_n)\cap N \text{ negative tests overall}\right),  
\end{equation}

where $P(\Tilde{K}= (K_0,...,K_n)\cap N \text{ negative tests overall})$ is the probability of testing all persons in one specific realization $\Tilde{K} = (K_0,...,K_n)$. As in the simple example, the probability for the draw have to be multiplied with the probability, that all infected people are tested negative. Using the equations~\eqref{equ:Hypergeom} and~\eqref{equ:negativTest}, we obtain

\begin{align} 
&P(\Tilde{K}= (K_0,...,K_n)\cap N \text{ negative tests overall}) \\
&= P(\Tilde{K}= (K_0,...,K_n))\cdot P(N \text{ negative tests overall}| K = \sum_{d=0}^{n}K_d \text{ persons are infected}) \\
&= \frac{\prod_{d=0}^{n}\binom{N_d}{K_d}}{\binom{M}{K}} \cdot \prod_{d=1}^n (1-s_{D_d})^{K_d} \\
&= \frac{\binom{N_0}{K_0}}{\binom{M}{K}} \cdot \prod_{d=1}^n \binom{N_d}{K_d}(1-s_{D_d})^{K_d}  
\end{align}

Thus, the likelihood function is modeled with the above equations and the posterior probability can be calculated via Bayes Theorem, as outlined in section~\ref{subsec:model}.

\section{Results and Discussions}
\label{section:results}

\subsection{Prior distribution for school classes}
\label{sec:schoolprior}

\begin{figure}[ht]
	\centering
	\includegraphics[width=\textwidth]{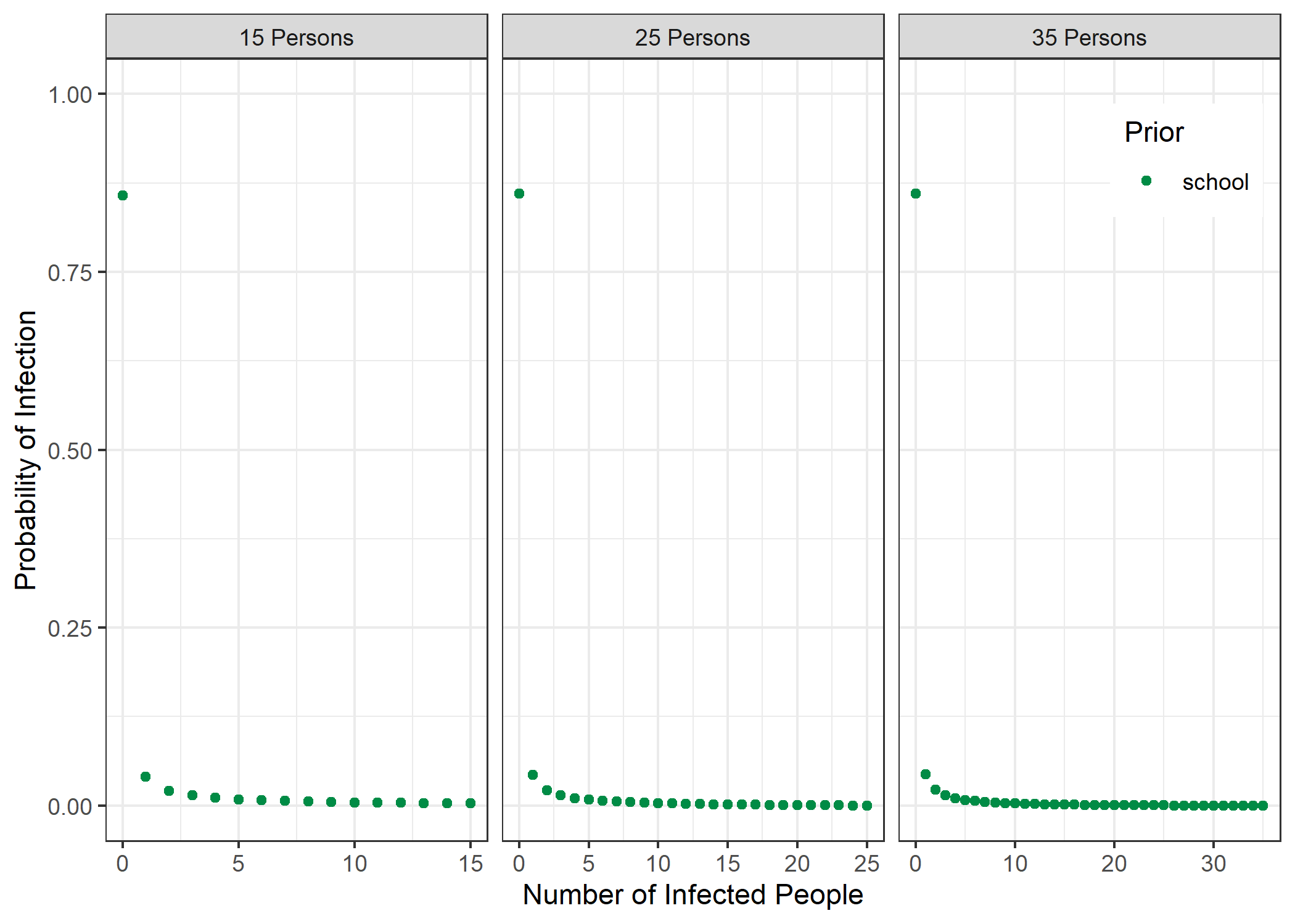}
	\caption{Plots of the Beta-binomial distribution as school class prior for $M = 15, 25, 435$}
	\label{fig-priorDistributions}
\end{figure}

We conducted a literature search about the probability of a SARS-CoV-2 outbreak in schools. This search was done with \textit{Google Scholar} with different combinations of the keywords \textit{SARS-CoV-2, COVID-19, transmission, school, pupils, education}. As a result, we found four studies that analyzed the transmission probability in school classes: a French study of one school~\cite{fontanet2020cluster}, an Irish study of all SARS-CoV-2 cases in school~\cite{heavey2020no}, an Australian study\cite{macartney2020transmission} analyzing different educational scenarios and a study in Luxembourg~\cite{mossong2021sars}.

In the Australian study, 15 school classes with at least one SARS-CoV-2 infected person were analyzed and for 3 school classes new SARS-CoV-2 infections were detected~\cite{macartney2020transmission}. No transmission of SARS-CoV-2 in 6 school classes was detected in the Irish study~\cite{heavey2020no}. In the French study, one outbreak was analyzed in detail, especially the transmissions of SARS-CoV-2 in one high school~\cite{fontanet2020cluster} that took place at a time prior to enforcement of strict pandemic hygiene rules. This study analyzed an outbreak at a whole school and not in one single school class.  The Study in Luxembourg analyzed the primary and secondary transmissions in school, however no information about transmission probability in school classes was reported~\cite{mossong2021sars}. Therefore, we decided not to the last two studies into account. Based on the two remaining sources, we obtain a first crude estimation of $P(K>0)$. Here, we considered all school classes equally. Then, the sum of all school classes with transmissions was calculated and divided by the sum of all considered school classes:

\begin{equation}
P(K > 0) = \frac{0+3}{6+15} =  \frac{3}{21} \approx 0.14.
\end{equation}

Moreover, the RKI recently published a report about SARS-CoV-2 outbreaks in different environments~\cite{rki2020Infektionsumfeld}. In this document an analysis about the average number of infected persons per outbreak was made. For school classes, an average number of $4.8$ infected persons was reported, which equals the conditional mean value:

\begin{equation}
E(K| K>0) = 4.8.
\end{equation}

With these two values, a unique Beta-binomial distribution can be determined as outlined in section~\ref{modeling:prior}. The distributions for school classes with $M = 15, 25, 35$ are plotted in Fig.~\ref{fig-priorDistributions}. For all cases, a suitable Beta-binomial distribution could be calculated. Small difference between the distributions can be only observed for the probability values with $K > 0$.

\subsection{Posterior probabilities and decision support for school classes}

\begin{figure}[ht]
	\centering
	\includegraphics[width=0.49\textwidth]{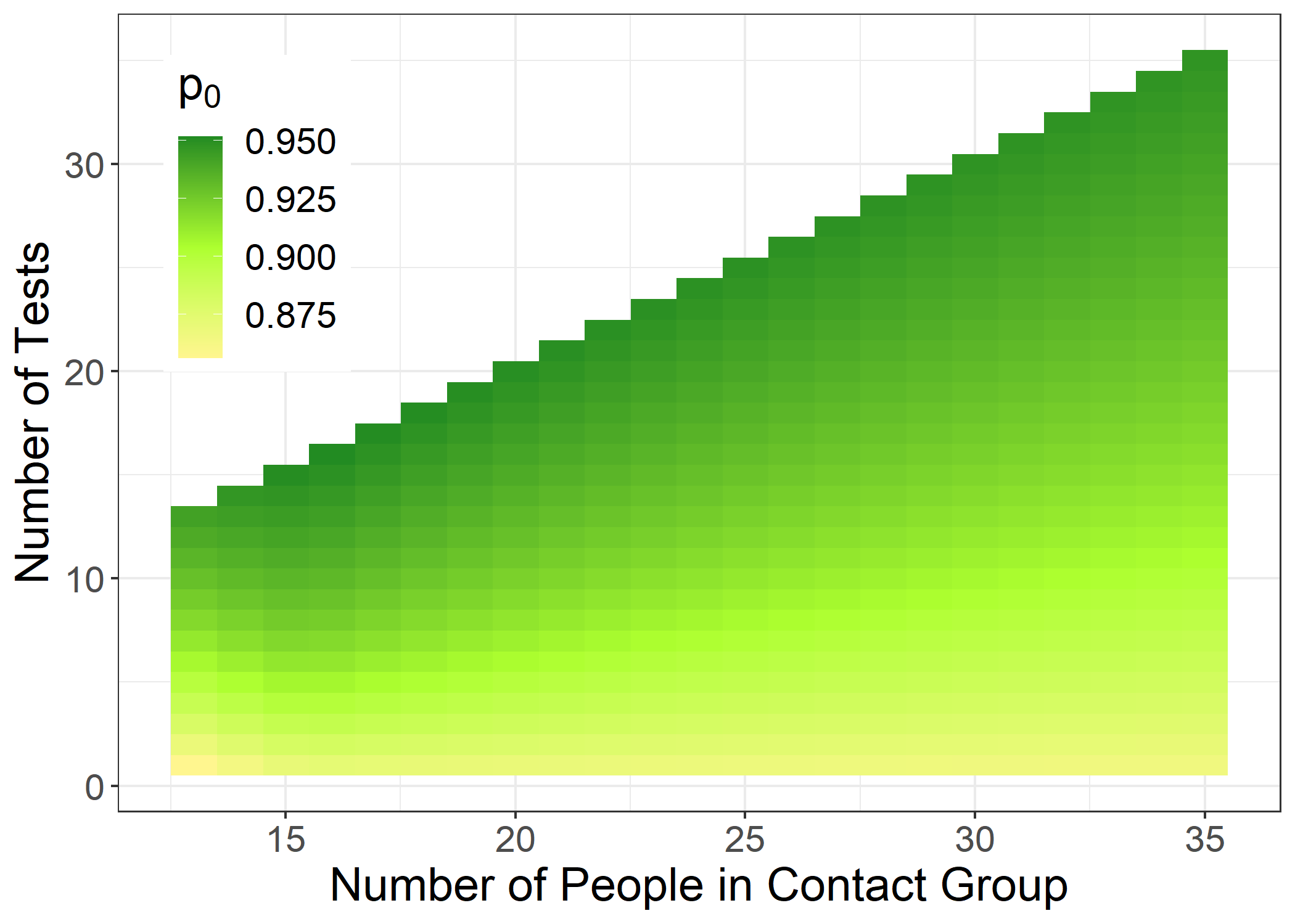}
	\includegraphics[width=0.49\textwidth]{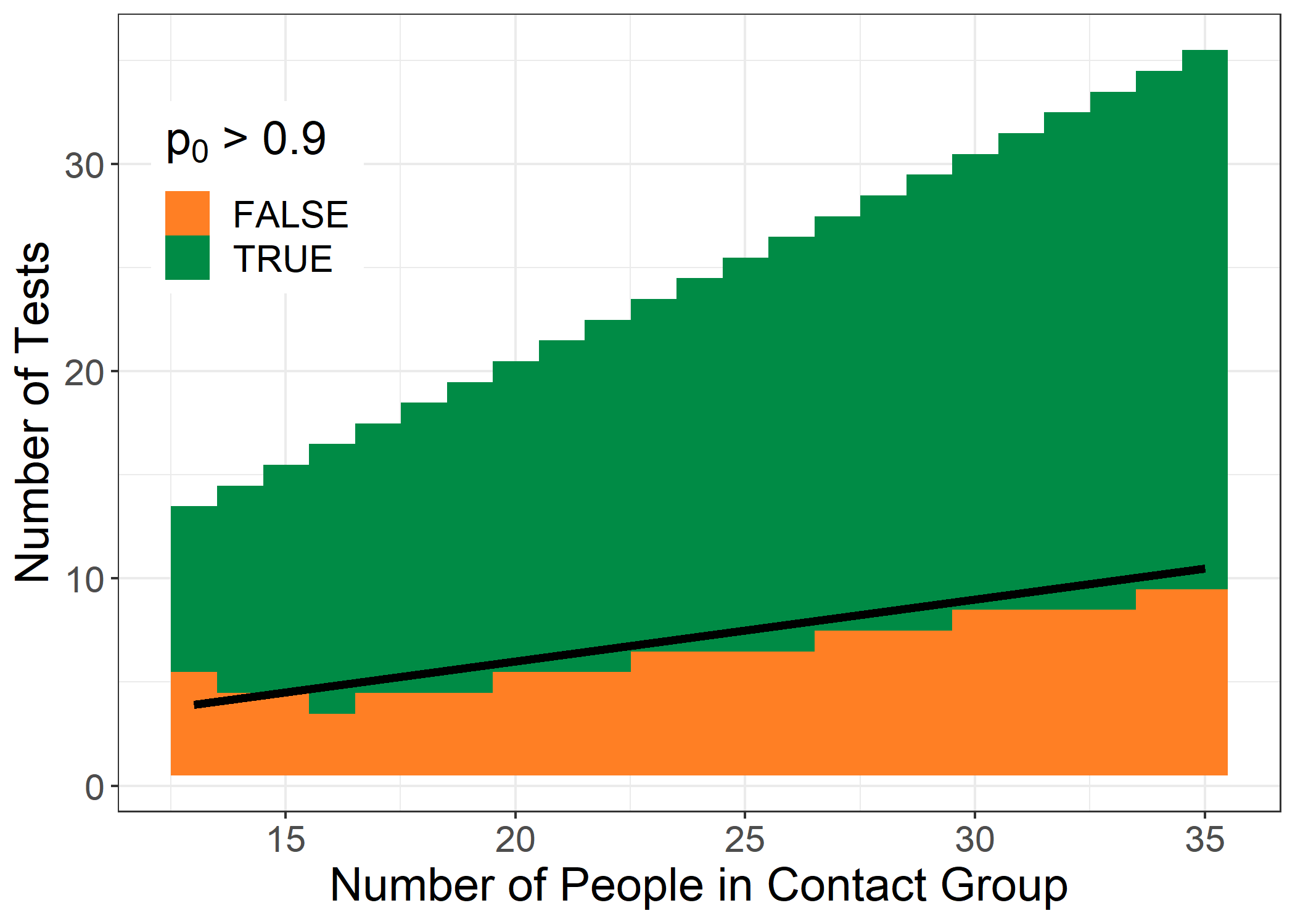}
	\caption{Left: Posterior probability that no transmission occurred given that $N$ out of $M$ persons are tested negative using the specific school prior for different group sizes and number of tests made 4 days after the group event. Right: Decision of a quarantine cancellation based the condition that $p_0 > 0.9$. The black line shows the situation that $30\%$ are tested.}
	\label{fig-posteriorDistributions}
\end{figure}

In this section, the school class scenario is further considered and the posterior probability is analyzed that no transmission occurred given that $N$ out of $M$ persons are tested negative. In Germany, school classes have usually at least 13 and not more than 35 pupils~\cite{Kultusminister2019}. Thus, we calculated this probability for different school class sizes $M \in \{13,...35\}$ and different number of tests $N$, which were all done 4 days after the group event. The results are shown in the left image of Fig.~\ref{fig-posteriorDistributions}. This probability ranges from $0.85$ up to $1.0$. One reason for this high minimum is the specific prior distribution, which should have the probability $0.86$ for no transmission. Thus, the posterior probability cannot be smaller than $0.86$ due to formula~\eqref{equ:Bayes}. Smaller values are only observed for $M < 15$, where no optimal parameters could be found with the optimization. Moreover, all determined posterior probabilities never reach $100\%$, because there is always some small remaining risk, that there could be some infected persons in the group, who were tested negative.


With this calculated probability, one can make a decision whether the quarantine can be canceled despite the fact that not all persons have been tested. Using for example the rule $p_0 > 0.9$, the quarantine will be canceled in most cases, as shown in the right image of Fig.~\ref{fig-posteriorDistributions}. Only when less than around $30\%$ of the group size have been tested, the result is $p_0$ is less than $90\%$ and the quarantine will be continued. 

\subsection{Decision about quarantine cancellation for school classes}

\begin{figure}[ht]
	\centering
	\includegraphics[width=0.72\textwidth]{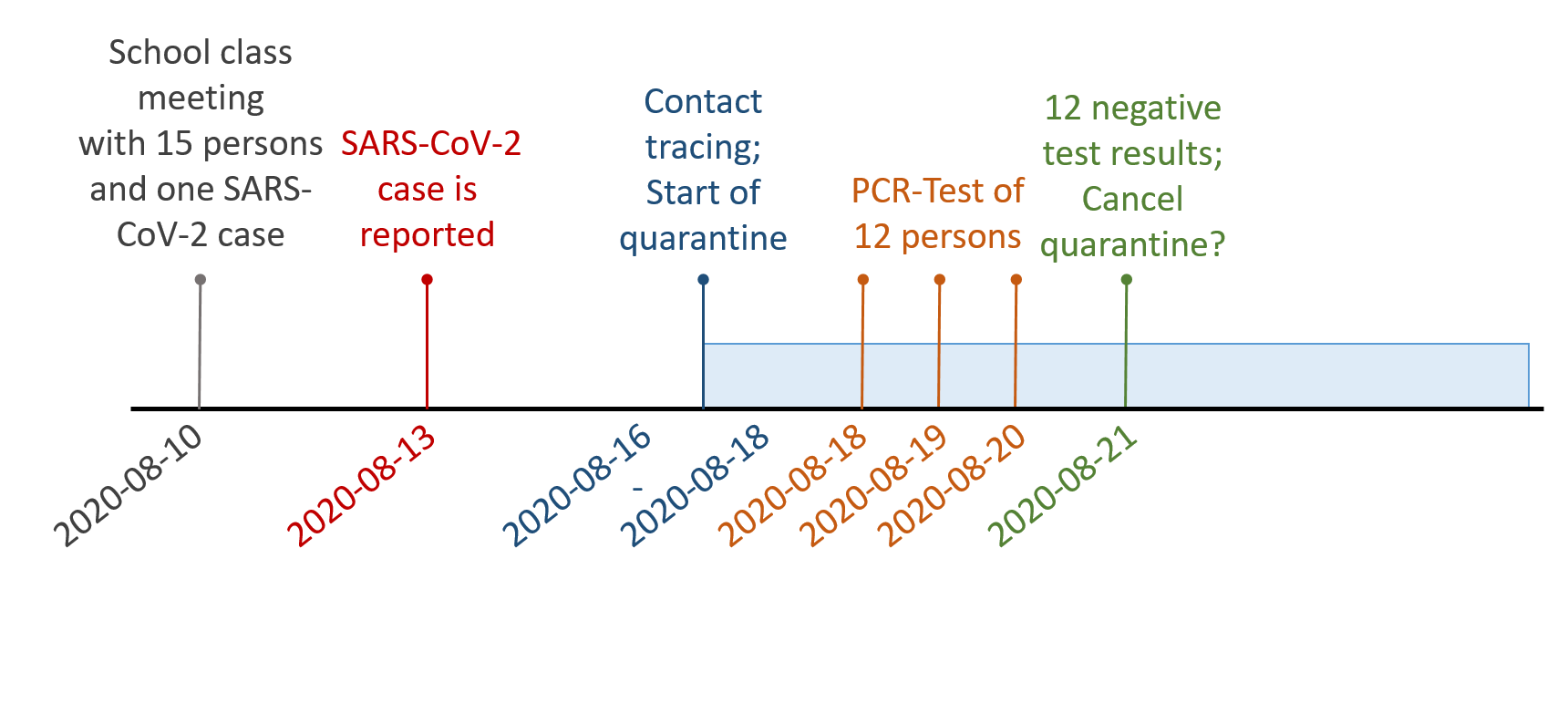}
	\caption{Timeline of the school class example.}
	\label{fig-timelineSchoolClass}
\end{figure}

The statistical model specified for school classes is tested and evaluated with one dedicated dataset provided by the health department Berlin-Reinickendorf. The dataset is shown in Table~\ref{tab:schoolClass}. We have 17 persons which were in contact with one infected person. All of them had contact on August $10^{\text{th}}$, for most of them it was the last contact. However, two of them have met the person later on until August $16^{\text{th}}$ and August $18^{\text{th}}$ again. These two persons do not fit into our group event setting and hence were excluded. Employees at the German health departments consider and decide about the quarantine of these persons separately on an individual basis.

As a result, we now have 15 persons with last contact on August $10^{\text{th}}$. 12 persons were tested and all test results are negative. One person was tested  on August $18^{\text{th}}$, ten persons on August $19^{\text{th}}$ and one person on August $20^{\text{th}}$. The timeline for this sequence of events is given in Fig.~\ref{fig-timelineSchoolClass}. With this information, we can calculate the posterior probability, that nobody at all was infected according to our model. 
With the prior distribution determined for a school class with $M = 15$ pupils, we obtain a probability of $p_0 = 98\%$, that no transmission of SARS-CoV-2 occurred in this setting. Based on the described decision rule of $p_0 > 0.9$, this result would support to cancel the quarantine for the whole group early. Indeed, no further SARS-CoV-2 cases were reported to the health department concerning this particular school class. Hence, the correct decision about cancellation of quarantine would have been made in this specific case.

\begin{table*}[t]
	\caption{A data set of a school class from the health department in Berlin-Reinickendorf, which contains the last contact date, the PCR test date and the PCR results.}
	\centering
	\begin{tabular*}{500pt}{@{\extracolsep\fill}ccccc@{\extracolsep\fill}}
		\toprule 
		\textbf{Case number} & \textbf{Date of last contact} & \textbf{PCR test date} & \textbf{PCR test result} \\
		\midrule
		1 & August 10, 2020 & --- & --- \\
		2 & August 10, 2020 & --- & --- \\
		3 & August 10, 2020 & --- & --- \\
		4 & August 10, 2020 & August 18, 2020 & negative \\
		5 & August 10, 2020 & August 19, 2020 & negative \\
		6 & August 10, 2020 & August 19, 2020 & negative \\
		7 & August 10, 2020 & August 19, 2020 & negative \\
		8 & August 10, 2020 & August 20, 2020 & negative \\
		9 & August 10, 2020 & August 19, 2020 & negative \\
		10 & August 10, 2020 & August 19, 2020 & negative \\
		11 & August 10, 2020 & August 19, 2020 & negative \\
		12 & August 10, 2020 & August 19, 2020 & negative \\
		13 & August 10, 2020 & August 19, 2020 & negative \\
		14 & August 10, 2020 & August 19, 2020 & negative \\
		15 & August 10, 2020 & August 19, 2020 & negative \\
		16 & August 16, 2020 & August 19, 2020 & negative \\
		17 & August 18, 2020 & August 24, 2020 & negative \\
		\bottomrule
	\end{tabular*}
	\label{tab:schoolClass}
\end{table*}

\section{Conclusions}
\label{section:discussions}

In this work, we introduced a statistical model for estimating the risk of an early group quarantine cancellation. Therefore, a day specific sensitivity and a specificity of $100\%$ of the PCR-test were considered for the likelihood distribution. Moreover, the Beta-binomial distribution was chosen as prior for the transmission of the SARS-CoV-2 and a method for determining its parameters for specific situations was proposed. This method was used for obtaining a prior distribution for school classes based on data obtained from literature. Based on the obtained prior distribution, we analyzed the usage of our statistical model for a decision support for quarantine cancellation. Offered as a web application or integrated in a software solution, our statistical model could be used by employees of German health departments and could facilitate their daily work. 

However, for the different probability distributions general and quite simplified assumptions were made. Several aspects like the number of initially infected persons in the group or other aspects of the group event (inside / outside, ventilation, activity type, interactions, ... ) were not addressed. Recently, a paper was published, which describes a model for estimating the aerosol transmission risk depending on several parameters e.g. room properties, event duration or number of participants~\cite{aerosol2020}. Based on this work, the probability that a transmission of SARS-CoV-2 occurred can be estimated. However, further information about the average number of infected persons are not available using this approach introduced in~\cite{aerosol2020}. Currently, there is still a lack of available data and published studies for extending our statistical model. In addition, for the likelihood we assumed that every contact made only one test. But normally contact persons are tested two times.

Since the prior specification is always subjective and depends on the availability of adequate evidence for model parameters, the sensitivity of the results to the prior should be addressed in the future. Different frameworks to approach robust Bayesian analysis have been described in the literature~\cite{berger2013statistical}. Furthermore, we assumed a random choice of tested persons, but in reality the contact persons with the highest risk would be chosen for tests. Thus, our model gives a conservative risk estimate in this case. The likelihood was specified for the PCR-tests, but this could be also adapted for other test types when the tests sensitivities and specificity are known.

In addition, we illustrated a decision rule based on the arbitrary threshold of $90\%$ for the posterior probability that no transmission occurred. However, it is not clear, if this is the best rule for making a decision. The threshold could be also set higher, e.g. $p_0 > 0.95$, to reduce the remaining risk. Future work should address the choice of the threshold in dependence of factors such as available test capacities. Moreover, we will implement other circumstances of the contact event (location, interactions, ...) in the model and will apply it for other specific scenarios. In a next step, the benefit of using our tool in routine cases has to be evaluated. For this purpose, an evaluation study with German health departments is planned.


\section*{Funding}
This research was funded by the Fraunhofer Internal Programs under Grant No. Anti-Corona 840242.

\section*{Acknowledgements}
We thank the team from the Fraunhofer ITWM, in particular Andreas Wagner and Stefanie Grimm, for the fruitful collaboration in the CorASiV project, especially for the collaborative development of the prototypical web application for German health departments. Furthermore, we thank Ann Christina Foldenauer (Fraunhofer IME - Institute for Molecular Biology and Applied Ecology, Frankfurt, Germany) for her valuable feedback on an earlier version of the manuscript. 

\section*{Author contributions:}
Sonja Jäckle: investigation, methodology, software, visualization, writing – original draft preparation, writing – review \& editing \\
Elias Röger: methodology, software, visualization, writing – review \& editing \\
Volker Dicken: investigation, methodology, writing – original draft preparation, writing – review \& editing  \\
Benjamin Geisler: investigation, writing – review \& editing \\
Jakob Schumacher: conceptualization, resources, writing – review \& editing \\
Max Westphal: methodology, investigation, supervision, writing – original draft preparation, writing – review \& editing \\

\section*{Conflict of interest}
The authors declare no potential conflict of interests.

\subsection*{Orcid}
\textit{Sonja Jäckle}: \href{https://orcid.org/0000-0002-2908-299X}{\includegraphics[width=0.02\textwidth]{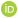}$\;$ https://orcid.org/0000-0002-2908-299X} \\
\textit{Elias Röger}: \\
\textit{Volker Dicken}: \href{https://orcid.org/0000-0003-1629-5811}{\includegraphics[width=0.02\textwidth]{orcid.png}$\;$ https://orcid.org/0000-0003-1629-5811} \\
\textit{Benjamin Geisler}: \\
\textit{Jakob Schumacher}: \href{https://orcid.org/0000-0002-9894-6085}{\includegraphics[width=0.02\textwidth]{orcid.png}$\;$ https://orcid.org/0000-0002-9894-6085} \\
\textit{Max Westphal}: \href{https://orcid.org/0000-0002-8488-758X}{\includegraphics[width=0.02\textwidth]{orcid.png}$\;$ https://orcid.org/0000-0002-8488-758X} \\

\bibliography{paper}

\end{document}